\documentclass[fleqn,twoside]{article}
\usepackage{espcrc2}

\usepackage{graphicx}
\usepackage[figuresright]{rotating}

\newcommand{\AmS}{{\protect\the\textfont2
  A\kern-.1667em\lower.5ex\hbox{M}\kern-.125emS}}

\title{Finite-size scaling and deconfinement transition in gauge theories}

\author{Roberto Fiore\address[CS]{Dipartimento di Fisica, Universit\`a della 
Calabria \\ \& Istituto Nazionale di Fisica Nucleare, Gruppo Collegato di Cosenza,
Italy},
Alessandro Papa\addressmark[CS]
and Paolo Provero\address{Dipartimento di Scienze e Tecnologie Avanzate, 
Universit\`a del Piemonte Orientale, Alessandria, \\
Dipartimento di Fisica Teorica, Universit\`a di Torino \\ 
\& Istituto Nazionale di Fisica Nucleare, Sezione di Torino, Italy}}
       
\begin{document}

\begin{abstract}

A new method is proposed for determining the critical indices of the deconfinement
transition in gauge theories, based on the finite-size scaling analysis
of simple lattice operators, such as the plaquette. A precise determination of 
the critical index $\nu$, in agreement with the prediction of the 
Svetitsky-Yaffe conjecture, is obtained for SU(3) gauge theory in 
(2+1)-dimension. Preliminary results for SU(2) in (3+1)-dimension are also given. 
\vspace{1pc}
\end{abstract}

\maketitle

\section{INTRODUCTION}

Given a $(d+1)$-dimensional pure gauge theory undergoing a continuous 
deconfinement transition at the critical temperature $T_c$, the order parameter 
of this transition is the Polyakov loop.
The $d$-dimensional effective model obtained by integrating out all degrees 
of freedom except the order parameter is globally invariant under the center
of the gauge group. This effective model possesses only short-range interactions
and therefore the phase transition, when continuous, is accompanied by long 
range fluctuations only in the order parameter~\cite{SY82}.

According to the Svetitsky-Yaffe conjecture~\cite{SY82}, the 
$(d+1)$-dimensional gauge theory and the $d$-dimensional effective model, if 
it also displays a second-order phase transition, belong to the same 
universality class. Therefore, their critical properties such as
critical indices, finite-size scaling, correlation functions at criticality, 
are predicted to coincide. The validity of the Svetitsky-Yaffe conjecture has 
been confirmed in several Monte Carlo analyses~\cite{test}. 

In the present work we have considered SU(3) in (2+1)-dimension
and SU(2) in (3+1)-dimension, which belong to the universality classes 
of the 3-state Potts model in 2-dimension and of the Z(2) (Ising) model 
in 3-dimension, respectively.

\section{FINITE-SIZE BEHAVIOR AT CRITICALITY}

For a $d$-dimensional statistical model, a way to extract critical 
indices is to study the finite-size behavior of suitable operators. 
If a mapping between operators in the $(d+1)$-dimensional gauge theory 
and operators in the $d$-dimensional effective model is established,
it is possible to exploit finite-size effects also in the gauge theory
to determine its critical indices. This would provide with a new method
for the computation of the critical indices of a pure gauge theory
with a second order phase transition.

The Svetitsky-Yaffe conjecture intrinsically establishes 
a correspondence between the Polyakov loop of the gauge theory and the 
order parameter of the effective model. 
The mapping of the plaquette operator of the gauge theory with 
a linear combination of the identity and the energy operators of the 
effective model has been shown in~\cite{GP97} for SU(2) in (2+1)-dimension,
which generates as effective model the 2-dimensional Ising model.
In this case, the critical behavior of the effective model is known exactly,
thanks to the methods of conformal field theory.
This mapping between operators has allowed to study, by means of
universality arguments, several non-perturbative features of gauge 
theories~\cite{FGP98-99}. 

In the present work the method is applied to the computation of the critical 
index $\nu$ of the correlation length. In the $d$-dimensional statistical model, the
finite-size behavior of the (lattice) energy operator $E$ is given by
\begin{equation}
\langle E\rangle_L\sim \langle E\rangle_\infty+k L^{\frac{1}{\nu}-d}\;,
\end{equation}
where $L$ is the lattice size, understood to be large enough,   
and $k$ is a non-universal constant. In the $(d+1)$-dimensional pure gauge 
theory, any operator $\hat{O}$, invariant under gauge and center symmetry
is expected to scale according to
\begin{equation} 
\hat{O}=c_I\ I + c_\epsilon\  \epsilon + \mbox{irrelevants}\;, 
\end{equation}
where $I$ is the identity operator and $\epsilon$ the (scaling) energy 
operator of the statistical model. Simple operators which satisfy
the mentioned symmetry requirements are Wilson loops and operators like
$ P(\vec x) P^{\dagger}(\vec y)$, where $P(\vec x)$ is the Polyakov loop 
at the site $\vec x$ and $\vec x$, $\vec y$ represent different sites 
of the $d$-dimensional space. This ansatz was introduced and tested in~\cite{GP97}
and used in~\cite{FGP98-99}.

\section{NUMERICAL RESULTS}

{\bf SU(3) in (2+1)-dimension\footnote{The results
presented in this subsection have been published in Ref.~\cite{FPP01}.}}

The lattice operators considered were the electric and
magnetic plaquette and $P(\vec x) P^{\dagger}(\vec y)$, with
$\vec x$ and $\vec y$ neighbor sites in the 2-dimensional spatial lattice.
All these observables have the required symmetry properties and can
be computed with high accuracy in Monte Carlo simulations.
Simulations were performed on lattices with temporal size $N_t=2$ and 
spatial sizes $N_x=N_y=L$ ranging from 7 to 30; $\beta$ was set at 
its critical value $\beta_c(N_t=2)=8.155$, taken from Ref.~\cite{EKLLLPS97}. 
The simulation algorithm was a mixture of one sweep of a 10-hit Metropolis 
and four sweeps of over-relaxation consisting in two updates of (random) 
SU(2) subgroups. For each simulation 400K equilibrium configurations were 
collected, separated each other by 10 updating steps. The error analysis 
was done by the jackknife method applied to data bins at different 
levels of blocking. The table of numerical results can be found 
in Ref.~\cite{FPP01}. 

\begin{figure}[htb]
\includegraphics[scale=0.37]{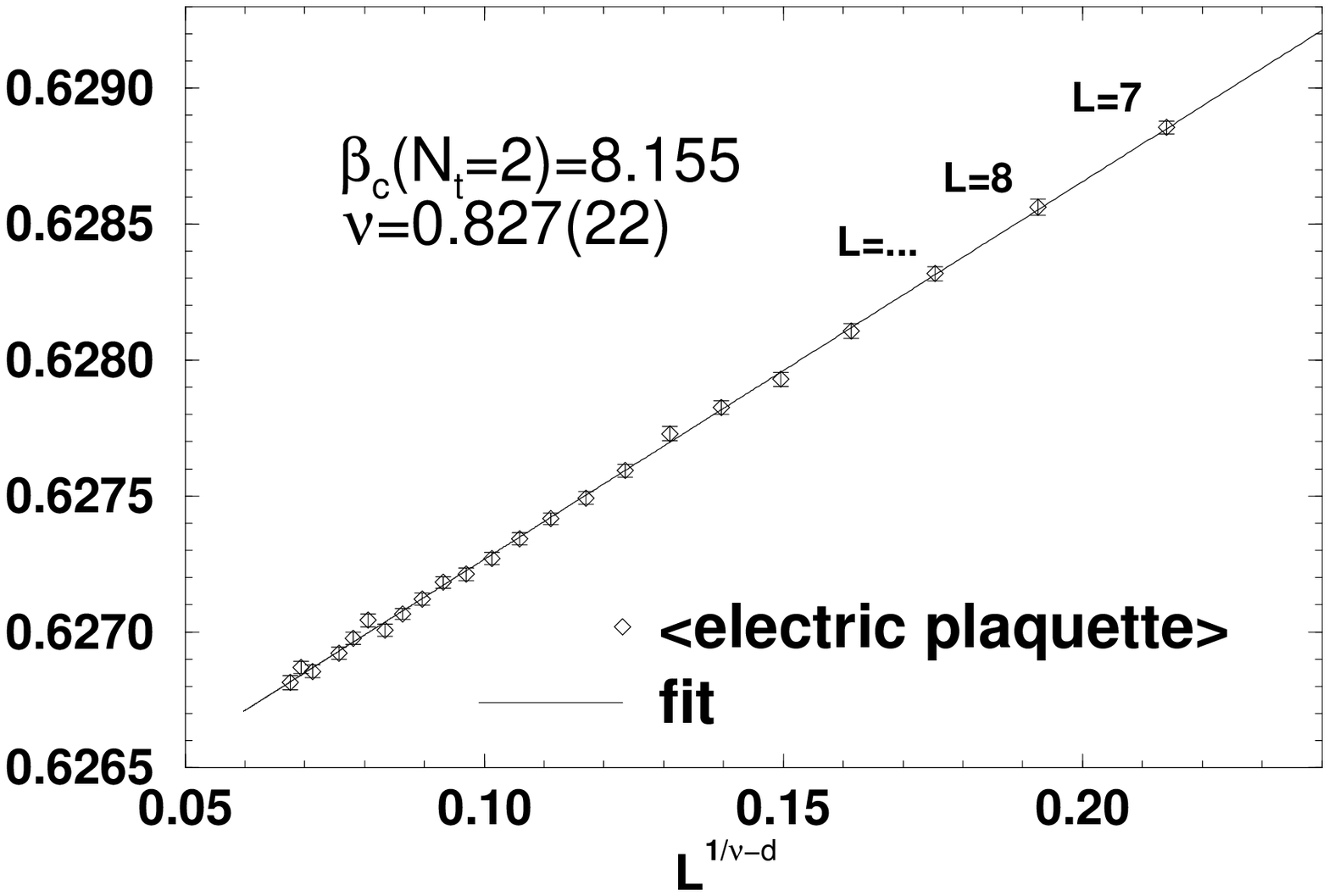}
\includegraphics[scale=0.37]{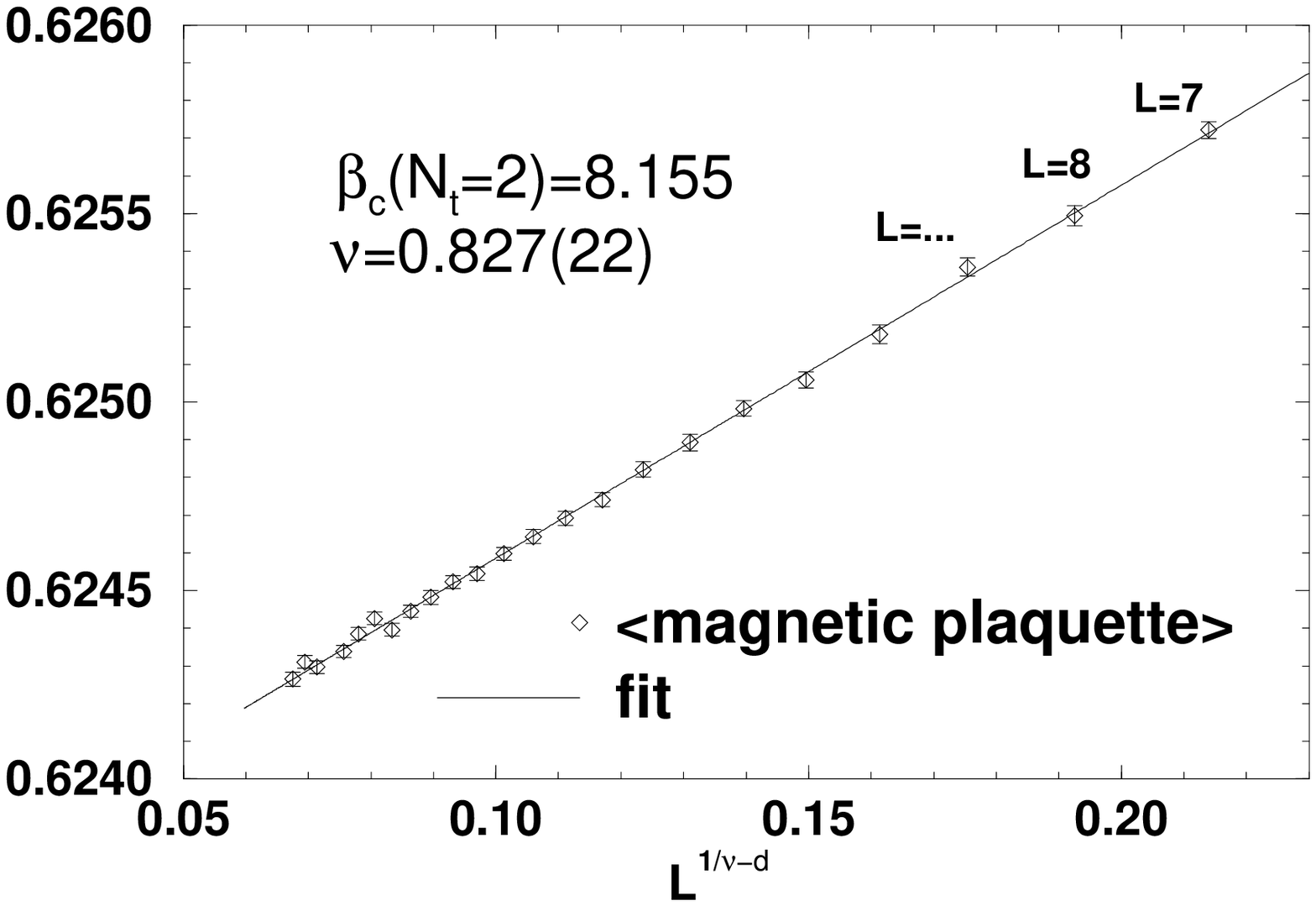}
\vspace{-0.8cm}
\caption{SU(3) in (2+1)-dimension: electric and magnetic plaquette 
{\it vs} $L^{1/\nu-d}$, where $\nu$ comes from the multibranched fit. 
Similar figure has been obtained for the Polyakov loop correlator.}
\label{fig_su3}
\end{figure}

A single multibranched fit was performed of the three data sets, taking
electric (magnetic) plaquettes from lattices with even (odd) $L$, in order 
to avoid cross correlations. Polyakov loop correlator data were measured 
separately and are therefore not correlated with the plaquette measurements. 
The result of the fit is $\nu=0.827(22)$, with $\chi^2_{\rm red}=0.84$ 
(see Fig.~\ref{fig_su3}), to be compared with the {\em exact} value in the 
2-dimensional 3-state Potts model, $\nu=5/6=0.833\cdots$. A 
previous Monte Carlo determination by the $\chi^2$ method in SU(3) in 
(2+1)-dimension~\cite{EKLLLPS97} gave $\nu_{MC}=0.90(20)$.

\vspace{0.5cm}
{\bf SU(2) in (3+1)-dimension\footnote{The following results 
are preliminary and have been obtained in collaboration with C.~Vena of the Dip. 
Fisica, Univ. Calabria.}}

The lattice operators considered are the electric and magnetic plaquette.
Monte Carlo simulations were performed on lattices with $N_t=2$  and 
$N_x=N_y=L$ ranging from 5 to 22, at the critical value of $\beta$, 
$\beta_c(N_t=2)=1.8735$, taken from Ref.~\cite{FKPS01}.
The simulation algorithm adopted is the over-relaxed heat-bath~\cite{PV91}
and the error analysis was done as in the previous case.
For each simulation we have collected so far a number of configurations ranging 
from 50K to 600K, according to the lattice size.

\begin{figure}[htb]
\includegraphics[scale=0.37]{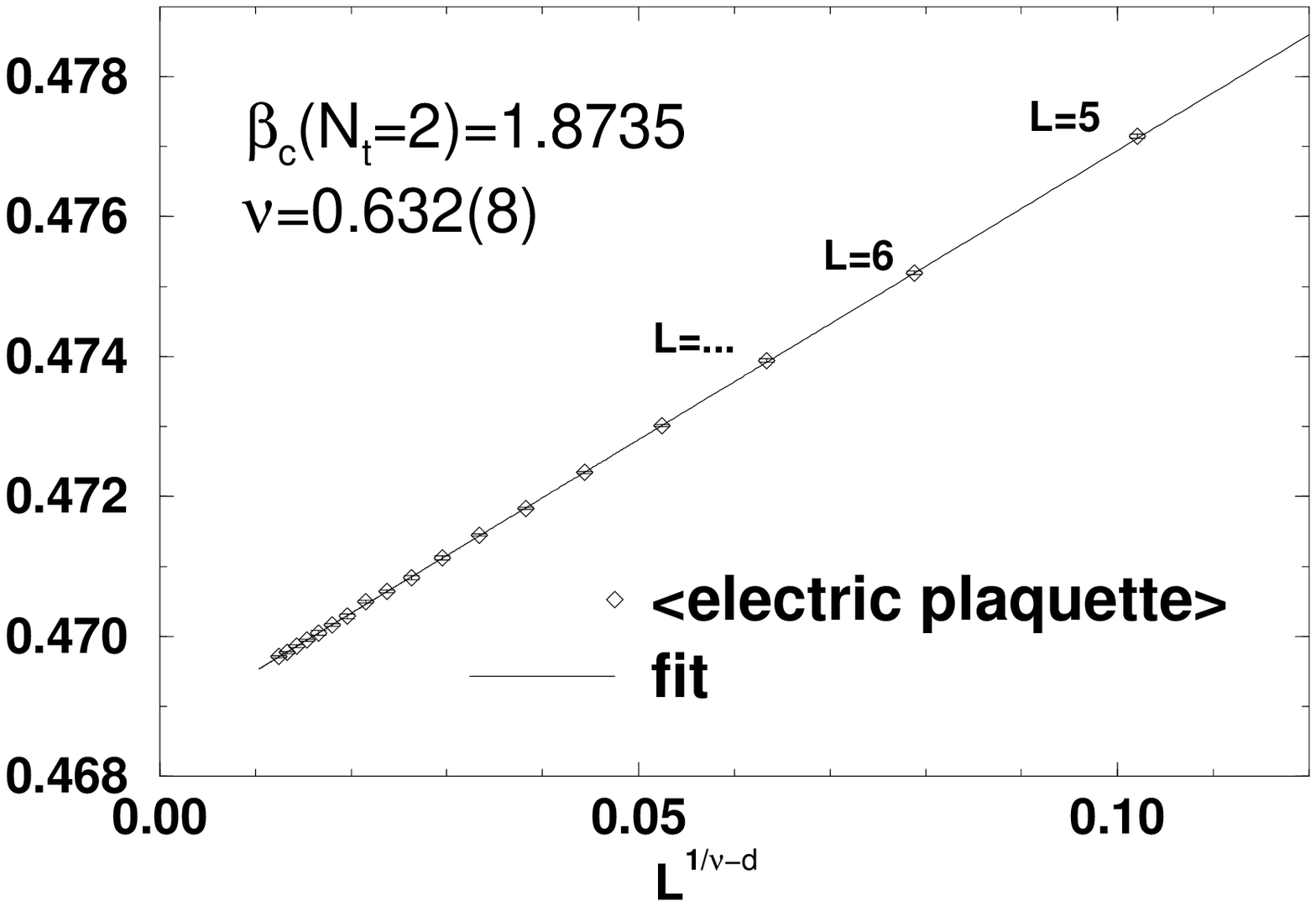}
\includegraphics[scale=0.37]{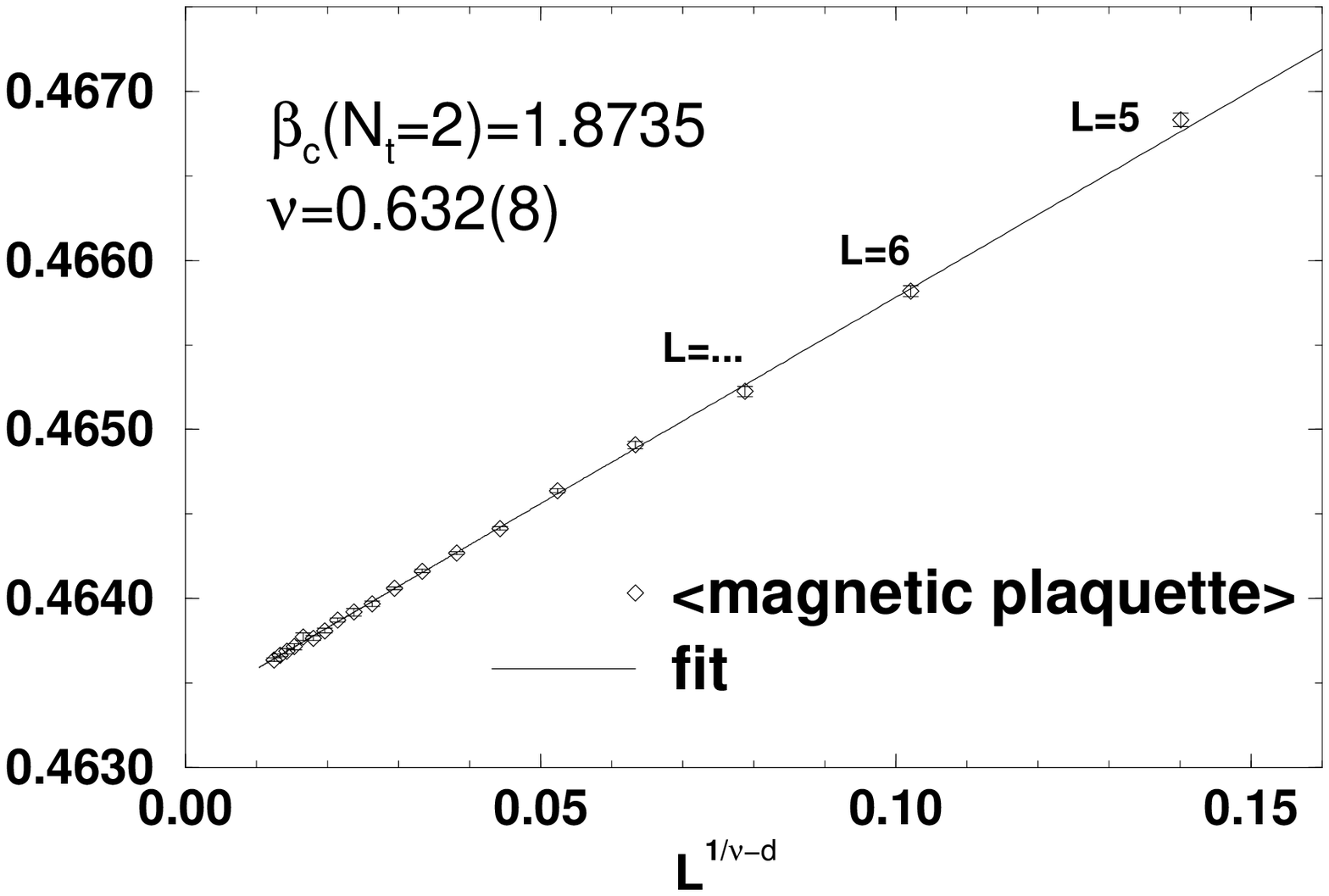}
\vspace{-0.8cm}
\caption{SU(2) in (3+1)-dimension: electric and magnetic plaquette 
{\it vs} $L^{1/\nu-d}$, where $\nu$ comes from the multibranched fit.}
\label{fig_su2}
\end{figure}

The multibranched fit on the electric (magnetic) plaquettes taken from lattices 
with even (odd) $L$ has given $\nu=0.632(8)$, with $ \chi^2_{\rm red}=0.30$
(see Fig.~\ref{fig_su2}). This result is to be compared with the determination 
of $\nu$ in the 3-dimensional Ising model. In this model, the high-temperature 
expansion method gave the following very accurate 
result~\cite{CPRV99}: $\nu=0.63002(23)$. Compare also with the Monte Carlo 
determination by 
the $\chi^2$ method in SU(2) in (3+1)-dimension~\cite{FKPS01,F01}: 
$\nu_{MC}=0.630(9)$.

\end{document}